\documentclass[
  reprint,
  amsmath,
  amssymb,
  superscriptaddress,
  aps,
  pre,
  10pt,
]{revtex4-2}

\usepackage{graphicx}       
\usepackage{dcolumn}        
\usepackage{bm}             
\usepackage{float}
\usepackage{amsmath}
\usepackage{amssymb}
\usepackage[colorlinks=true, allcolors=blue]{hyperref}

\newcommand{\ihat}{\hat{\textbf{i}}}
\newcommand{\jhat}{\hat{\textbf{j}}}
\newcommand{\khat}{\hat{\textbf{k}}}

\newcommand{\phihat}{\hat{\textbf{$\phi$}}}
\newcommand{\infrac}[2]{
  \mathchoice
  {#1/#2} 
  {#1/#2} 
  {#1/#2} 
  {#1/#2} 
}

\begin{document}

\title{Controlling \texorpdfstring{$\gamma$}{gamma}-photon production with additively manufactured microstructured targets}

\author{J. Luoma}
\affiliation{School of Applied \& Engineering Physics, Cornell University}
\author{A. Kemp}
\affiliation{Lawrence Livermore National Laboratory}
\author{T. G. White}
\affiliation{Department of Physics, University of Nevada, Reno}
\author{D. Rusby}
\affiliation{Lawrence Livermore National Laboratory}
\author{H. Chen}
\affiliation{Lawrence Livermore National Laboratory}
\author{G. Shvets}
\affiliation{School of Applied \& Engineering Physics, Cornell University}

\begin{abstract}
Nonlinear inverse Compton scattering (NICS) in high-density targets offers a promising route for generating high flux, broadband sources of MeV $\gamma$-photons which are important for applications like radiography, photonuclear physics, and pair production. Microstructured targets enable control over the NICS process to help shape the photon energy, angular distribution, and total yield. This work presents particle-in-cell (PIC) simulations exploring target microstructures compatible with two-photon polymerization (TPP) additive manufacturing, which is a promising technique that allows high-volume and reproducible printing of sub-wavelength features as small as 200~nm. We investigate how microstructures influence NICS radiation and merge simulation analysis with simple analytical models that guide target design. These results establish pathways to optimize targets for low-divergence, high-energy photon beams with $2.0\cdot10^{8}$ photons/J exceeding 100 MeV and targets for high yields delivering upwards of $2.0\cdot10^{11}$ photons/J above 1 MeV in the multi-PW regime.
\end{abstract}

\maketitle

\section{Introduction}

Broadband $\gamma$-photon sources in the MeV energy range offer great utility to high-field, high-energy-density (HED), and nuclear science. MeV photons can drive nuclear excitations, electron-positron pair production, and other effects of interest in the strong-field regime \cite{chen2023perspectives,meuren2016semiclassical,zilges2022photonuclear}. In HED science, $\gamma$-rays are valuable in radiography for constraining areal density, shock velocity, and material opacity in targets otherwise inaccessible to optical probes \cite{barbato2019quantitative,wood2018ultrafast,balcazar2025multi}. Mechanisms for generating hard photons with lasers include betatron, bremsstrahlung, and both linear and nonlinear inverse Compton scattering (ICS \& NICS). Betatron emission is routinely co-produced with wakefield-accelerated electrons, but high spectral yields in the MeV-regime remain difficult to achieve. Bremsstrahlung offers larger yields and is naturally broadband, but source control is limited by electron transport in large converters of size $\mathcal{O}\approx1$~mm \cite{wegert2024demonstrating}.

ICS offers a pathway for generating high-energy $\gamma$-photons by upscattering laser photons with relativistic electrons. Typically, ICS is associated with low conversion efficiency; however, recent experiments using near-critical foams have achieved ICS conversion efficiencies of order $10^{-3}$ and yields of approximately $10^{10}$ photons/J above 10 keV \cite{shou2023brilliant}. Pushing this concept into the strongly nonlinear regime, where the laser normalized vector potential $a_0\gg1$, enables multi-photon absorption and a field-dependent emission rate \cite{di2012extremely}. Simulation studies of near-critical targets demonstrate that NICS can further enhance efficiency and average photon energy, making this process an excellent next-generation x-ray source \cite{Formenti_2022_Bremsstrahlung_double_layer,formenti2024three,galbiati2023numerical}.

High-density NICS schemes typically require the laser to (1)~accelerate electrons to relativistic energy and (2)~supply photons for scattering. The electron acceleration requirement makes NICS difficult in simple foil targets which motivates the addition of structures to the foil front surface to increase efficiency. Aerogel and carbon-nanotube foams offer a route to higher efficiency by providing a near-critical density plasma that strongly couples to the laser \cite{galbiati2026numerical,maffini2026nanofoam}. However, the chemical processes used to create such foams produce random density patterns in the laser focal volume, often requiring target homogenization with laser pre-pulse \cite{shou2023brilliant}. Consequently, the exact plasma conditions become difficult to diagnose and translate to different laser systems. 

This work explores microstructures that are compatible with modern two-photon polymerization (TPP) techniques for additive manufacturing of targets. State-of-art TPP can achieve feature sizes as small as 200~nm and programmable laser-based printing ensures target reproducibility. Recent advances in parallelization methods allow high-volume production of microstructured targets, which can be leveraged for high-repetition rate experiments \cite{gu20253d}. TPP targets have also seen recent success in high-efficiency ion acceleration experiments \cite{tochitsky2025high}. The combination of sub-wavelength features, customized geometries, and mass production motivates the development and optimization of $\gamma$-photon sources based on TPP methods, as is done in this work.

\section{Foam target design and simulation}

\begin{figure}[t]
    \centering
    \includegraphics[width=\columnwidth]{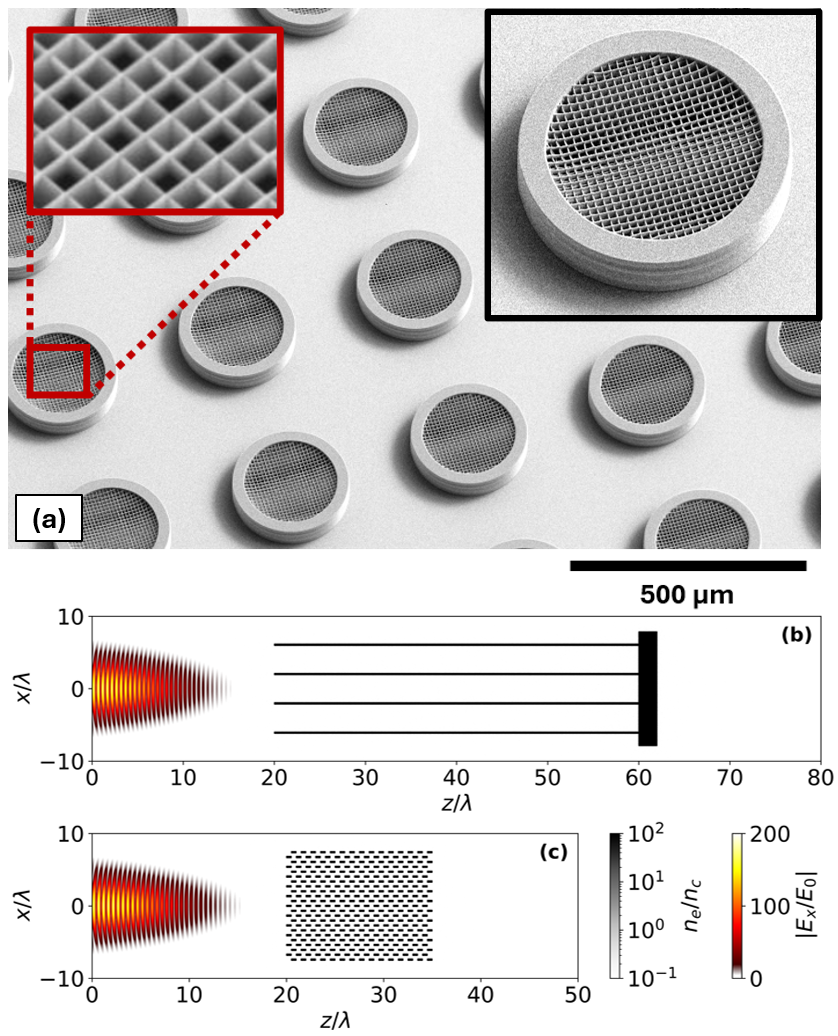}
    \caption{Modern TPP printing methods allow parallel production of microstructured targets, as shown in the (a) scanning electron microscope image of staggered-filament foams (image courtesy of Xiaoxing Xia). Panels (b) and (c) show TPP targets simulated in 2D PIC, with the laser entering from the left boundary, in (b) a channel array pattern and (c) a staggered-filament foam pattern.}
    \label{fig:foam_targets}
\end{figure}

This work presents foam designs, shown in Fig.~\ref{fig:foam_targets}, that are compatible with modern TPP techniques for 3D-printing of microstructures. We study two concepts: (A)~a channel-reflector target for producing high-energy photons and (B)~a staggered-filament foam for high photon yields. The channel arrays consist of continuous filaments with a $250$~nm wall thickness and spacing $s=4.05$~\textmu m. The staggered-filament foam has a spacing $s=1.35$~\textmu m and staggers the filaments every half-wavelength ($\lambda=1$~\textmu m) to increase surface area. The filament thickness along $x$ is taken to be $250$~nm which is near the limit of current TPP methods \cite{gu20253d}.

The channel scheme creates a miniature electron accelerator that collimates a high-energy electron beam towards an overcritical, reflecting boundary. When the laser reaches the reflector, NICS photons are generated in a bright $\gamma$-flash. In contrast, the staggered foam utilizes crossed filaments to promote rapid laser absorption. Self-focusing and magnetic field generation promote scattering without the need for a reflecting boundary. Increasing the foam fill-fraction allows the laser to couple to a high electron density, enhancing the photon yield.

Foam targets are investigated with 2D particle-in-cell (PIC) simulations using SMILEI \cite{derouillat2018smilei}. The target is modeled as solid plastic with $n_e=300n_c$, 300 particles/cell, and cell dimensions of 20~nm by 20~nm. The laser parameters represent a multi-PW beam with $a_{0}=200$, $\lambda=1$~\textmu m, $\tau_{p}=30$~fs, and $\mathcal{E}_{L}=200$~J polarized along $x$ \cite{radier202210}. The high intensity necessitates modeling the quantum regime of NICS where stochastic emission of photons has a non-negligible impact on electron dynamics due to radiation reaction \cite{mackenroth2014quantum,niel2021classical}. Radiation is modeled using the Monte Carlo method in SMILEI \cite{duclous2010monte,lobet2016modeling}.

\begin{figure}[t]
    \centering
    \includegraphics[width=\columnwidth]{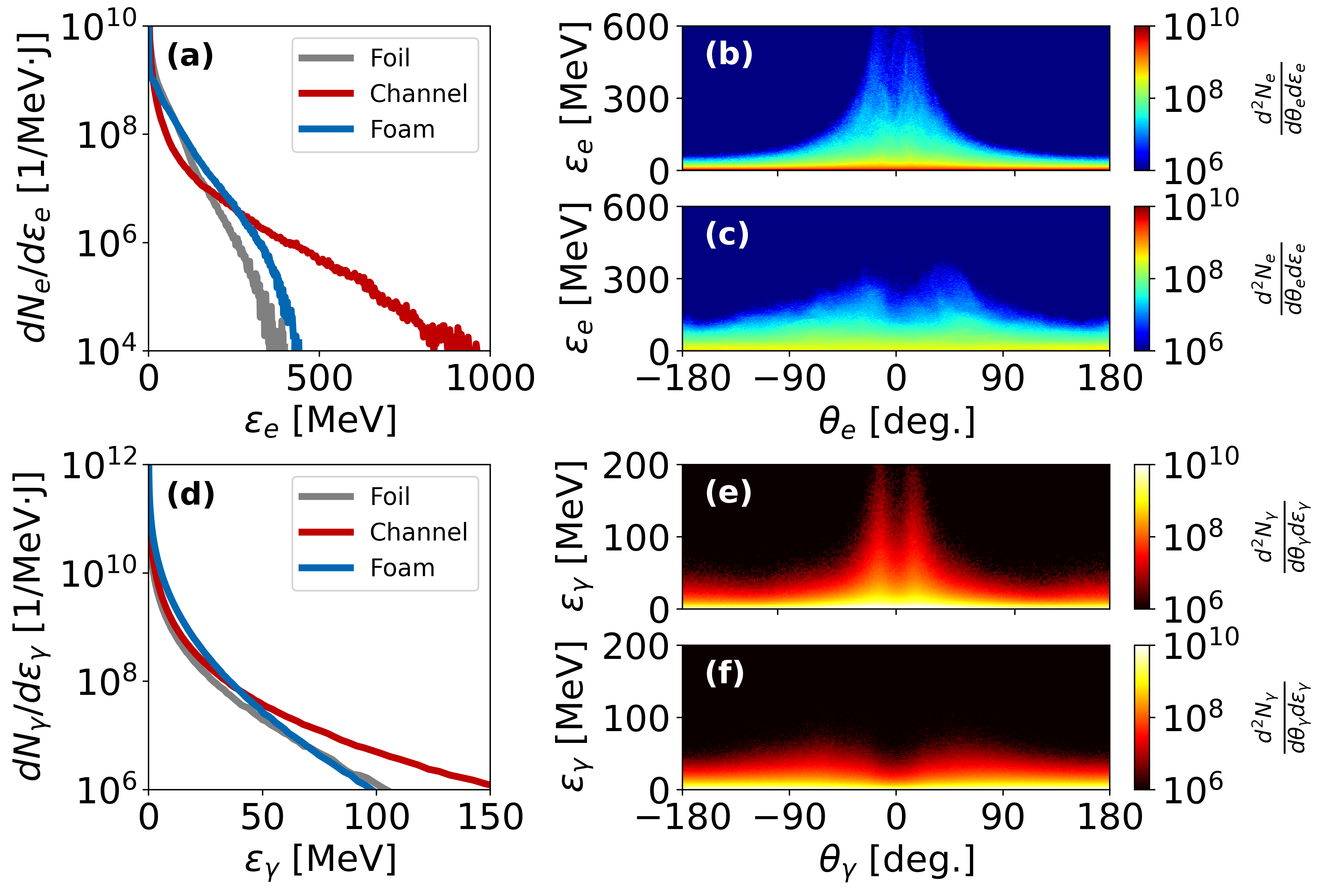}
    \caption{Energy spectra for (a) electrons and (d) photons are provided for channel array, staggered-filament foams, and foil targets. The angular distribution of electrons is provided in panel (b) for the channel array and (c) for the staggered foam. Similarly, the angular distribution of photons is n panel (e) for the channel array and (f) for the staggered foam.}
    \label{fig:foam-electrons-and-photons}
\end{figure}

The foam microstructure influences the average energy and total charge of accelerated electrons, which in turn determines the NICS photon distribution. Fig.~\ref{fig:foam-electrons-and-photons} summarizes the overall performance of $\gamma$-photon emission for the channel array, staggered foam, and a foil target serving as a control. Electrons, presented in panels (a)-(c), gain higher energies in the channels due to a longer acceleration length-scale. The foam, in contrast, absorbs the entire laser in a short distance ($\approx10\lambda$) and couples to a high-density electron beam. Self-focusing effects also produce a wide-divergence electron source, whereas the walls of the channel array provide a collimating effect.

The time-integrated $\gamma$-photon distribution, represented in Fig.~\ref{fig:foam-electrons-and-photons}(d)-(f), mirrors the spectral and angular characteristics of the radiating electrons \cite{pouyez2024multiplicity,pouyez2025kinetic}. The radiation patterns giving rise to the final photon distribution can be isolated temporally in PIC simulations and presented as a set of distinct signatures that emerge from the laser-target interaction. The key physics of interest can then be reconstructed analytically by solving for radiative power $P$ for a simplified electron distribution and electromagnetic field configuration.

\section{Theory}
Nonlinear inverse Compton scattering leverages strong fields to coherently scatter multiple laser photons in a single interaction event. Compared with the linear regime, NICS enhances conversion efficiency of laser energy into $\gamma$-photons with yields reaching and exceeding 10\% J/J foreseeable with new and upcoming multi-PW laser facilities \cite{nakamura2012high}.

\begin{figure*}[t]
    \centering
    \includegraphics[width=\textwidth]{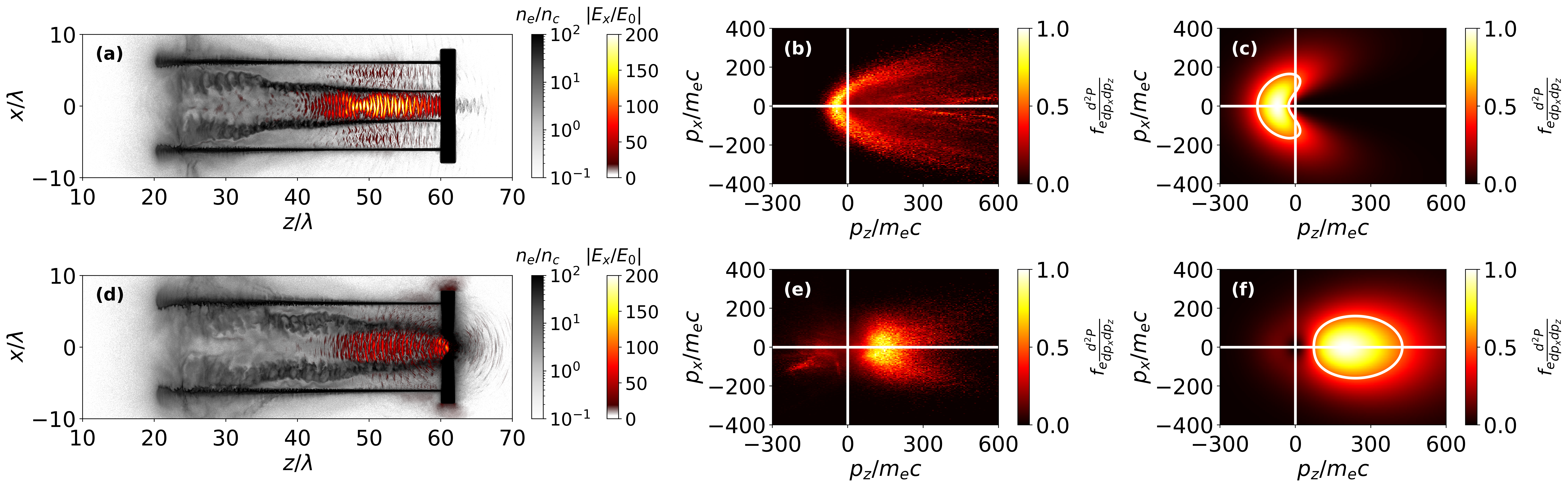}
    \caption{Electron radiative power $P(\chi_e)$ exhibits unique patterns during (a) laser propagation at $t=240$~fs and (d) formation of a standing wave at $t=290$~fs in a channel array. The laser field and electron density in the channel are shown in orange and grey colormaps, respectively. In $p_x\text{-}p_z$ momentum space, an arc-like pattern forms from thermal electrons counter-propagating against the laser, with (b) showing PIC data and (c) an analytical model using Eqs.~\ref{eqn:Ex-field-free}-\ref{eqn:By-field-free}. A contour is provided at half-maximum power. The standing wave shifts radiative power to the forward-traveling electron beam as evident by (e)~PIC data and (f) analytical modeling with Eqs.~\ref{eqn:Ex-field-standing}-\ref{eqn:By-field-standing}.}
    \label{fig:electron-rad-channel}
\end{figure*}

Achieving high scattering rates is contingent on electrons experiencing fields comparable to the Schwinger critical field $E_{cr}=m_e^2 c^3/e\hslash\approx 1.3\cdot10^{18}$~V/m \cite{schwinger1948quantum}. Constants $m_e$, $c$, $e$, and $\hbar$ are electron mass, speed of light, fundamental charge, and the reduced Planck's constant. While current lasers fall far below the $E_{cr}$ threshold, the electric field in the electron rest frame is boosted by the Lorentz factor $\gamma_{e}$ \cite{jackson_classical,zangwill2012modern}. This effect is quantified by the Lorentz-invariant quantum parameter $\chi_e$ \cite{ritus1985quantum},

\begin{equation}
    \chi_e = \frac{\varepsilon_e}{E_{cr}m_ec^2}\sqrt{(\vec{E}+\vec{v}_{e} \times \vec{B})^2-(\vec{v}_{e} \cdot \vec{E})^2/c^2},
    \label{eqn:quantum_parameter}
\end{equation}

for an electron of velocity $\vec{v}_{e}$ and energy $\varepsilon_e=\gamma_em_ec^2$ in an arbitrary electric $\vec{E}$ and magnetic $\vec{B}$ field. Similarly, the quantum parameter for a $\gamma$-photon of energy $\varepsilon_\gamma$ takes a similar form,

\begin{equation}
    \chi_\gamma = \frac{\varepsilon_{\gamma}}{E_{cr}m_ec^2}\sqrt{(\vec{E}+\vec{n} \times c\vec{B})^2-(\vec{n} \cdot \vec{E})^2},
    \label{eqn:quantum_parameter_ph}
\end{equation}

where $\vec{n}$ is a unit vector for the photon direction of travel. The NICS emission rate is typically solved in the locally constant field approximation (LCFA), where the laser is treated as a uniform, coherent state over the quantum formation length of the photon. LCFA is reasonable provided the formation length of photon emission is much smaller than the laser wavelength. This condition is expressed as $(\chi_e/a_0^3)(\varepsilon_e-\varepsilon_\gamma)/\varepsilon_\gamma\ll1$ and is satisfied for the near-MeV regime of interest in this work \cite{di2018implementing}. Under this approximation, the spectral power of NICS is expressed as \cite{niel2021classical}

\begin{equation}
    \frac{dP}{d\varepsilon_\gamma}=\frac{2\alpha m_e^2 c^4}{3\hbar}\frac{1}{\varepsilon_e}G\left(\chi_e,\chi_\gamma\right),
    \label{eqn:differential_power}
\end{equation}


\begin{equation}
    \begin{split}
        G(\chi_e,\chi_{\gamma})&=\frac{3\sqrt{3}}{4\pi}\frac{\chi_{\gamma}^2}{\chi_e}\nu K_{2/3}(\nu) \\
        &+\frac{\sqrt{3}}{2\pi}\frac{\chi_{\gamma}}{\chi_e}\int_{\nu}^{\infty}K_{5/3}(y)dy,
    \end{split}
    \label{eqn:quantum_emissivity}
\end{equation}


where $\nu=2\chi_\gamma/3\chi_e\left(\chi_e-\chi_\gamma\right)$ and $G(\chi_e,\chi_{\gamma})$ is the quantum emissivity. $K_n$ is a modified Bessel function of the second kind. This formulation ensures energy conservation by enforcing a vanishing probability that an electron of energy $\varepsilon_e$ emits a photon energy $\varepsilon_{\gamma}>\varepsilon_e$ \cite{niel2021classical}.

Integrating over all photon energies provides the average radiated power emitted by an electron, which is useful for benchmarking PIC simulations against simplified analytical models,

\begin{equation}
    P\left(\chi_e\right) = \frac{2\alpha m_e^2 c^4}{3\hbar} \chi_e^2 g(\chi_e),
    \label{eqn:radiated_power}
\end{equation}

\begin{equation}
    g(\chi_e) = \int_{0}^{\chi_e}\frac{1}{\chi_e^3}G\left(\chi_e,\chi_{\gamma}\right) d\chi_{\gamma},
    \label{eqn:Gaunt}
\end{equation}

where $g(\chi_e)$ is interpreted as the Gaunt factor for NICS and suppresses the $\chi_e^2$ scaling for large $\chi_e$. In the limit $\chi_e\xrightarrow{}0$, the radiated power reduces to the classical Larmor power $P_{Larmor}=2\alpha m_e^2 c^4 \chi_e^2/3\hbar$.

\begin{figure*}[t]
    \centering
    \includegraphics[width=\textwidth]{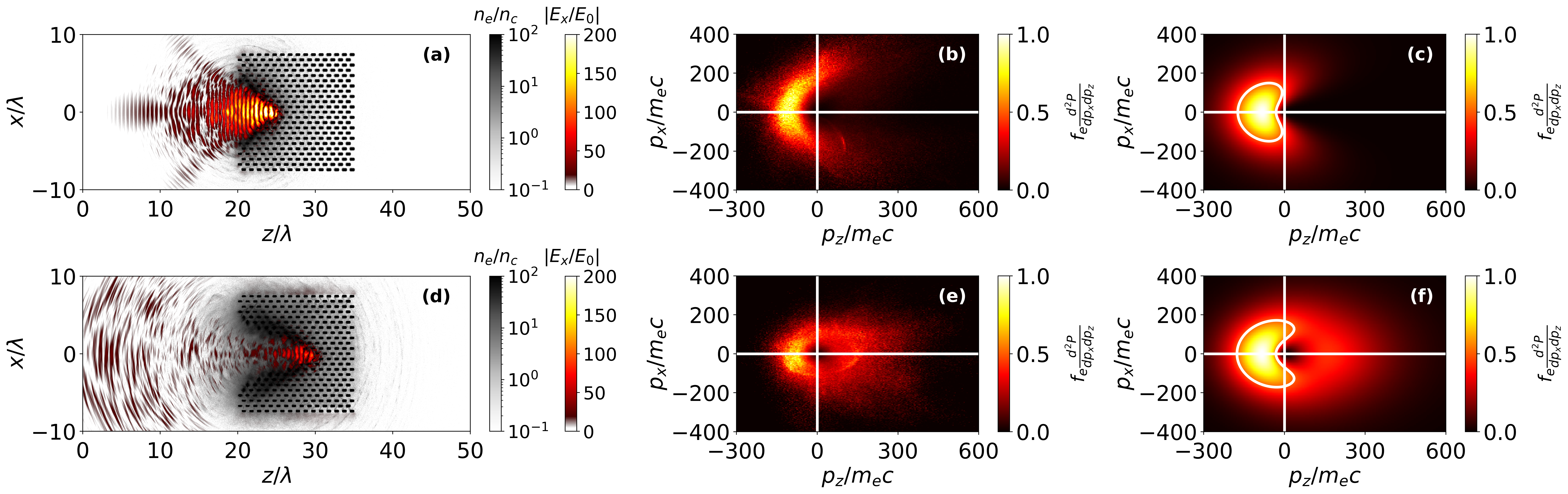}
    \caption{Electron radiative power $P(\chi_e)$ in a foam target at (a) $t=150$~fs and (d) $t=190$~fs with the laser field in orange and electron density in grey. During early stages of propagation, radiative power in $p_x\text{-}p_z$ momentum space forms an arc-like pattern evident in (b) PIC data and (c) analytical modeling with Eqs.~\ref{eqn:Ex-field-free}-\ref{eqn:By-field-free}. The contour shows half-maximum power. As self-focusing effects strengthen, strong azimuthal magnetic fields and longitudinal electric fields drive ring-like radiative patterns seen in (e) PIC data and (f) the analytical model with the inclusion of strong-focusing effects using Eqs.~\ref{eqn:Ex-field-focusing}-\ref{eqn:Ez-field-focusing}.
    }
    \label{fig:electron-rad-staggered}
\end{figure*}

\section{Results}
\subsection{Radiation in channel arrays} 

Channel array microstructures offer an accelerator-reflector approach to NICS photon generation. When the laser enters the channel, the field extracts electrons from the walls and generates a beam through direct laser acceleration (DLA) \cite{valenta2024direct}. Wall expansion fills the channel with a low density, hot plasma. Fast electrons form attosecond bunches co-propagating with the laser and scatter photons reflected at the solid-density boundary at the end of the channel \cite{naumova2004attosecond}. The electron momentum distribution is approximated by a drifting relativistic Maxwell--Jüttner distribution \cite{wright1975relativistic,bret2010exact},

\begin{equation}
f_e(\boldsymbol{u_e}) =\frac{1}{4\pi\Gamma^{2}\Xi K_{2}\left(\Gamma^{-1}\Xi^{-1}\right)}e^{-\left(\gamma_e(\boldsymbol{u_e})-\boldsymbol{\beta}\cdot\boldsymbol{u_e}\right)/\Xi},
\label{eqn:Maxwell-Juttner}
\end{equation}

where $\boldsymbol{u}_e=\boldsymbol{p_e}/m_ec$ is normalized momentum, $\gamma_e(\boldsymbol{u_e})=\sqrt{1+u_e^2}$, $\boldsymbol{\beta}=\langle\boldsymbol{v_e}\rangle/c$ is the normalized mean electron velocity, $\Gamma=(1-\beta^2)^{-1/2}$, and $\Xi=k_BT_e/(m_ec^2)$ is the normalized effective electron-temperature parameter. Here, $K_2$ is a modified Bessel function of the second kind. The distribution satisfies $\int f_e(\boldsymbol{u_e})\,\mathrm{d}^3u_e=1$. Fits to the PIC electron distribution give $\boldsymbol{\beta}=0.5\hat{\boldsymbol{k}}$ and $\Xi\approx60$.

The channel length $L_{0}$ and width $s$ influence the final energy and charge of the electron beam as well as the quality of the $\gamma$-flash. Here we present $\infrac{L_{0}}{\lambda}=40$ and $\infrac{s}{w_{0}}=1.5$ to obtain a high-energy electron beam, with a cutoff near 1 GeV, while ensuring sufficient laser energy reaches the reflector. If $L_{0}$ is too large or $s$ too narrow, attenuation depletes the laser resulting in a weak $\gamma$-flash. The channel structure also contributes to the collimation of electrons, resulting in a beam with a root-mean-square (rms) divergence of $\theta_{rms}=26.6^{\circ}$ above 100 MeV.

Electron density in the channel during laser propagation remains low enough to avoid aggressive self-focusing, and the laser maintains a field strength comparable to vacuum-intensity during the accelerator stage. The field configuration during propagation is reasonably approximated as a plane wave with a Gaussian envelope \cite{esarey2009physics},

\begin{equation}
    h_{\pm}(r,\xi_{\pm})=e^{-r^2/w_0^2}e^{-2\ln{2}(\xi_{\pm}/c\tau_p)^2},
    \label{eqn:envelope}
\end{equation}

and

\begin{equation}
    \vec{E}_x=\Re\left(E_{0}h_{+}e^{i\Phi_{+}}\right)\ihat,
    \label{eqn:Ex-field-free}
\end{equation}

\begin{equation}
    \vec{B}_y=\Re\left(\frac{1}{c}E_{0}h_{+}e^{i\Phi_{+}}\right)\jhat.
    \label{eqn:By-field-free}
\end{equation}

Here $\Phi_{\pm}=\pm k\zeta-\omega_{L}\tau$ and $\xi_{\pm}=c\tau\mp\zeta$ with $\tau=t-t_r$ and $\zeta=z-z_r$ defined about the reflection point at $(t_r, z_r)$. The laser frequency is $\omega_{L}=2\pi c/\lambda$ and $E_0=a_0m_ec\omega_{L}/e$. Co-propagation of the laser and electrons suppresses $\chi_e$ and $P(\chi_e)$. Consequently, the electron beam produces minimal radiation during the accelerator stage in channel array foams. However, thermal plasma electrons produce a nearly isotropic distribution in momentum space ($p_x$ vs. $p_z$) which can radiate and produce $\gamma$-rays. Scattering is favored for thermal electrons traveling anti-parallel with respect to the laser and, to a lesser extent, electrons with large transverse momentum ($p_x$).

The co-propagating radiative mode gives rise to distinct arc-like patterns when $P(\chi_e)$ is projected in momentum space, as demonstrated in Fig.~\ref{fig:electron-rad-channel}(a)-(c). Panel (a)  shows the location of the laser along with the surrounding plasma electron density at the timestep of interest. Panels (b) and (c) provide histograms for $P(\chi_e)$ in electron momentum space for PIC data and analytical integration, respectively. The contour in (c) corresponds to half-max power and agreement between the analytical and simulated descriptions of $P(\chi_e)$, weighted by $f_e$, confirms that our simplified picture of the laser-target interaction is reasonable.

When the laser reaches the reflector, a standing wave forms with a field strength roughly twice the incident field. The transverse electric field shifts phase by $e^{i\pi}$ resulting in a new field configuration,

\begin{equation}
    \vec{E}_x=\Re\left(E_{0}h_{+}e^{i\Phi_{+}} - E_{0}h_{-}e^{i\Phi_{-}}\right)\ihat,
    \label{eqn:Ex-field-standing}
\end{equation}

\begin{equation}
    \vec{B}_y=\Re\left(\frac{1}{c}E_{0}h_{+}e^{i\Phi_{+}} + \frac{1}{c}E_{0}h_{-}e^{i\Phi_{-}}\right)\jhat.
    \label{eqn:By-field-standing}
\end{equation}

During the reflection, the electron beam experiences a sharp increase in $\chi_e$ and a corresponding boost in radiative power. The $P(\chi_e)$ pattern of the electrons shifts to high longitudinal momentum $p_z$, as indicated in Fig.~\ref{fig:electron-rad-channel}(d)-(f). Correspondingly, the radiation produced by the high-energy beam dominates over the thermal electron population, resulting in a bright $\gamma$-flash. Photons produced in the flash inherit the angular divergence of the electron beam, producing a collimated beam of high-energy photons with rms divergence of $\theta_{rms}=19.8^\circ$ for photons above 100 MeV containing $1.5\cdot10^8$ photons/J with $\theta\le\theta_{rms}$. Low-energy photons produced by both thermal and beaming electrons are comparatively isotropic. The final photon distribution is the cumulation of both radiative modes creating a broadband spectrum with $8.3\cdot10^{10}$ photons/J ($\eta_{\gamma}=8.4\%$) above 1 MeV and $2.0\cdot10^8$ photons/J ($\eta_{\gamma}=0.45\%$) above 100 MeV.

\begin{figure}[t]
    \centering
    \includegraphics[width=\columnwidth]{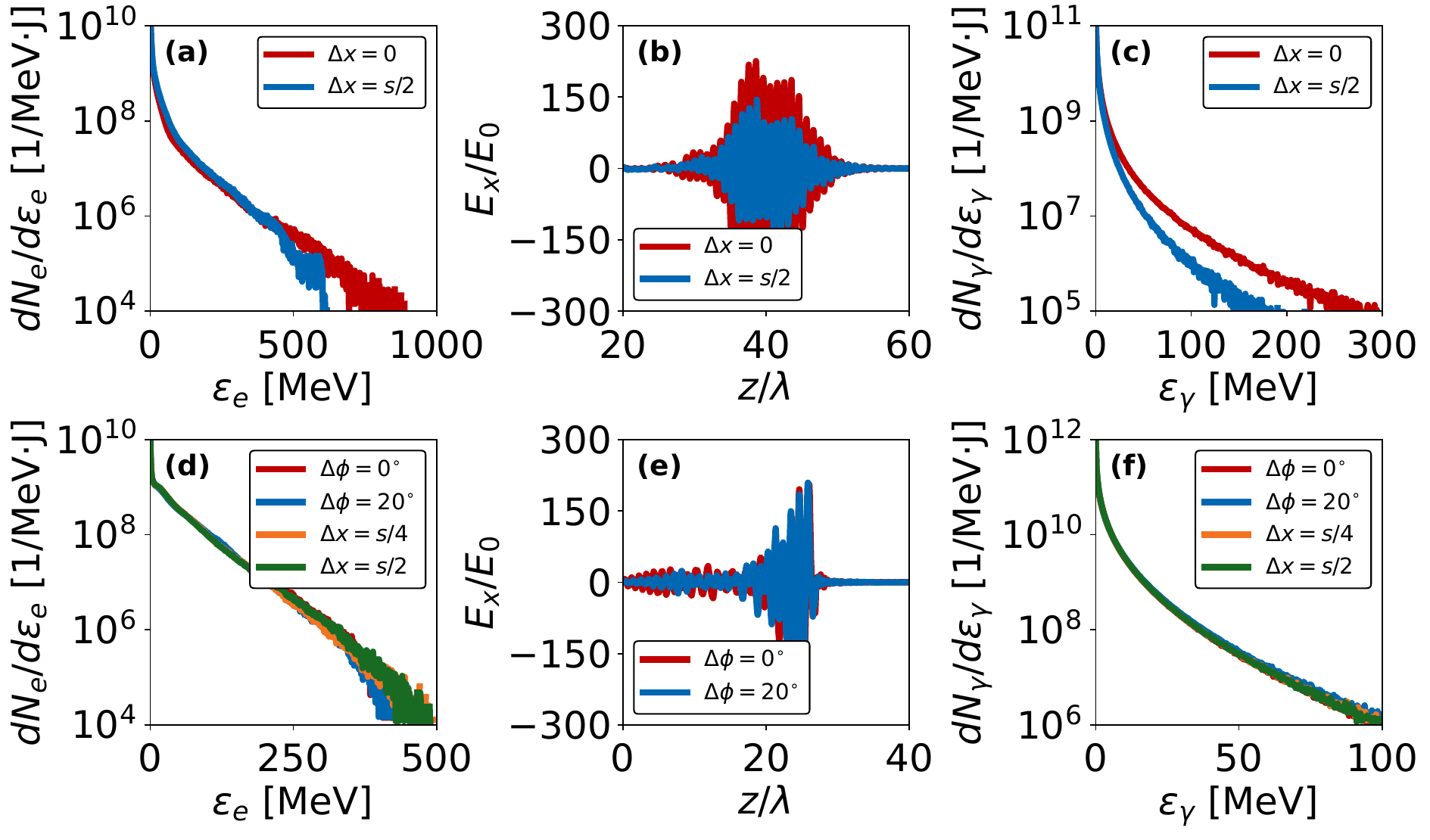}
    \caption{Robustness of NICS schemes with (a)-(c) channel arrays and (d)-(f) foams is tested by varying laser pointing with $\Delta x$. In the case of foam targets, target steering is explored with $\Delta\phi$ indicating the angle between target normal and the laser axis.
    }
    \label{fig:robustness}
\end{figure}

\subsection{Radiation in dense foams}

A staggered-filament foam enhances surface area to promote homogenization during the laser-target interaction. Filaments expand rapidly under the strong fields of the laser to create a plasma with an average density controlled by the foam fill fraction. The foam presented in Fig.~\ref{fig:foam_targets} has an average electron density of $n_e=55n_c$ below the relativistic-critical limit $\gamma n_c\approx140n_c$, where $n_c=\epsilon_{0}m_{e}\omega_{l}^2/q_{e}^2$ is the classical critical density. Constants $\epsilon_{0}$, $m_{e}$, $q_{e}$ are vacuum permittivity, electron mass, and electron charge. $\omega_{l}$ is the laser frequency.

The high-density foam promotes strong coupling between the laser and electrons to yield high-charge beams. The dense plasma also results in aggressive self-focusing of the laser, producing a highly scattered population of hot electrons. Based on PIC data, $\vec{\beta}=0.3\khat$ and $\Xi\approx60$ are used for the relativistic Maxwell-Jüttner distribution.

Laser self-focusing generates strong longitudinal electric fields $E_z$ that can be estimated from the paraxial approximation \cite{davis1979theory}. Strong electron currents create an azimuthal magnetic field $B_{\phi}$ which contributes to NICS radiation. Cumulatively, the field configuration in a tight focus takes the form,

\begin{equation}
    \vec{E}_x=\Re\left(E_{x}h_{+}e^{i\Phi_{+}}\right)\ihat,
    \label{eqn:Ex-field-focusing}
\end{equation}

\begin{equation}
    \vec{B}=\Re\left(\frac{1}{c}E_{x}h_{+}e^{i\Phi_{+}}\right)\jhat+\frac{r}{w_0}e^{-r^2/w_0^2}B_{\phi}\phihat,
    \label{eqn:Bazm-field-focusing}
\end{equation}

\begin{equation}
    \vec{E}_z=\Re\left(-i\frac{2x}{kw_0^2}E_{x}h_{+}e^{i\Phi_{+}}\right)\khat.
    \label{eqn:Ez-field-focusing}
\end{equation}

In early stages of the laser-target interaction, self-focusing effects are small and $P(\chi_e)$ produces arc-like patterns in momentum space as seen in Fig.~\ref{fig:electron-rad-staggered}(a)-(c). As discussed previously, field configurations similar to plane waves suppress radiation in the laser-forward direction. Strong self-focusing, however, introduces large $B_{\phi}$ and $E_z$ fields that interact with larger regions of the electron momentum space. Consequently, $P(\chi_e)$ produces ring-like patterns as observed in Fig.~\ref{fig:electron-rad-staggered}(d)-(f).

The foam target produces a wide-divergence photon distribution with $\theta_{rms}=110^\circ$ and $80^\circ$ above 1 and 50 MeV, respectively. The total yield of $2.0\cdot10^{11}$ photons/J above 1 MeV is significantly larger than the equivalent yield using the channel array, increasing the conversion efficiency to $\eta_{\gamma}=15\%$. However, the channel array delivers higher efficiency for photons beyond 50 MeV. For comparison, the foam yield above 100 MeV is only $1.7\cdot10^{7}$ photons/J or $\eta_{\gamma}\ll0.1\%$ compared to the channel array yield $\eta_{\gamma}=0.45\%$. The choice of microstructure is then contingent on application-specific requirements, depending on the desired photon energy and flux.

\section{Robustness and scaling of photon yield}

\begin{figure}[t]
    \centering
    \includegraphics[width=\columnwidth]{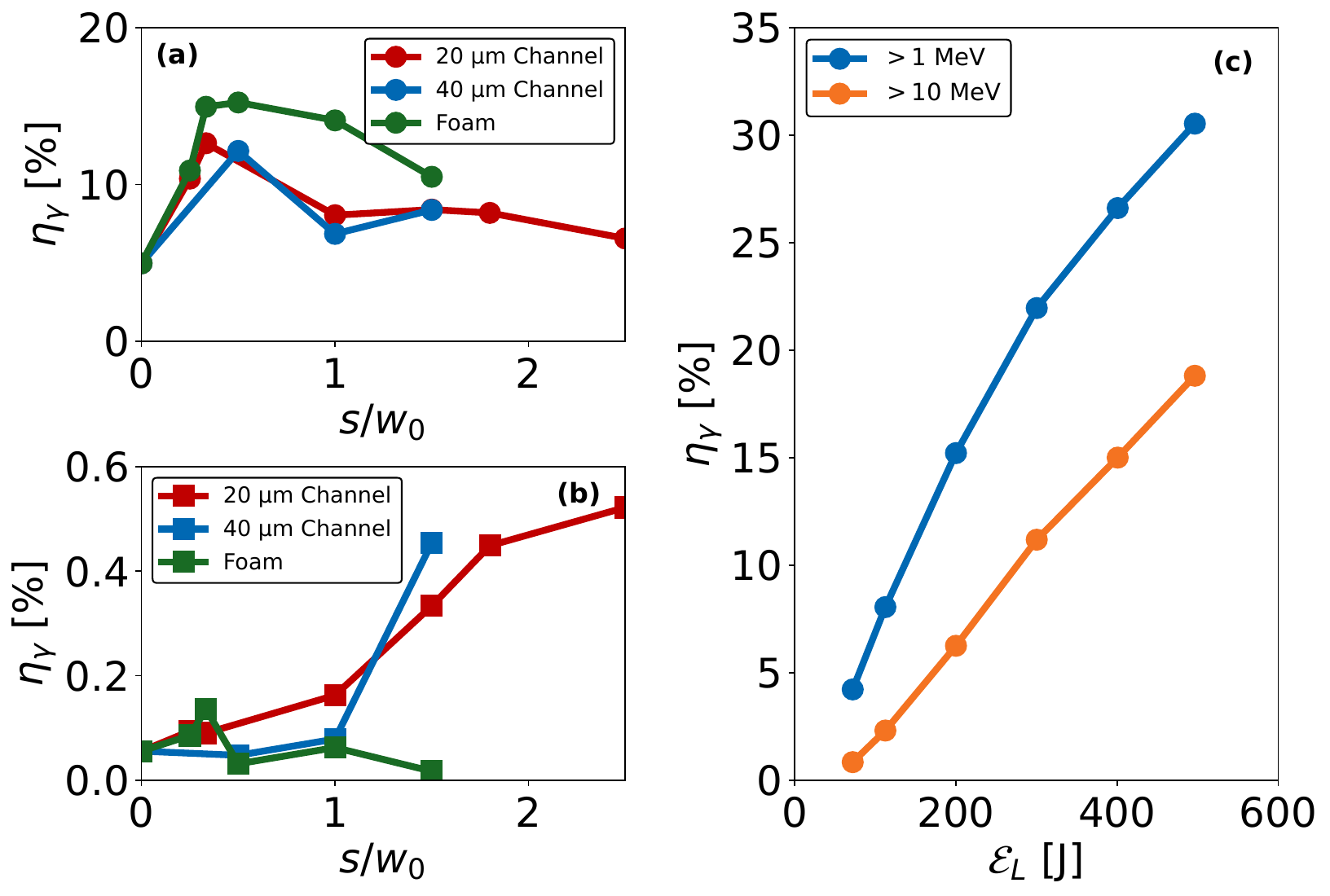}
    \caption{The scaling of $\gamma$-photon conversion efficiency with filament spacing $s$ for photons (a)~$\varepsilon_{\gamma}>~1$~MeV and (b)~$\varepsilon_{\gamma}>100$~MeV. A (c)~scan for staggered-filament foams over laser energy for a fixed $\tau_p=30$~fs and $w_0=2.7$~\textmu m.}
    \label{fig:photons-scaling}
\end{figure}

Laser pointing instabilities are inevitable when operating high power lasers with tight focusing geometry. The robustness of NICS generation in microstructures serves as another important metric when designing targets, as illustrated in Fig.~\ref{fig:robustness}. Channel arrays designed for the $\gamma$-flash regime are the most prone to shot-to-shot variations due to diffraction effects introduced by the finite wall thickness. We investigate this by shifting the laser pointing by $\Delta x=s/2$ and comparing scenarios where the laser is either centered on-channel or centered on-wall. In the on-wall case, the electron acceleration suffers at the highest electron energies but otherwise provides a similar electron temperature and dose. However, diffraction drops the field strength inside the channel roughly two-fold as shown in Fig.~\ref{fig:robustness}(b). Since $P(\chi_e)$ scales approximately as $\chi_e^2$, the overall NICS performance decreases nearly four-fold as corroborated by Fig.~\ref{fig:robustness}(c).

In contrast, the foam relies on average density effects where the target fill-fraction controls dynamics to leading order. The staggered filament pattern greatly benefits robustness by ensuring the laser sees approximately the same target conditions regardless of pointing. Fig.~\ref{fig:robustness}(d)-(f) demonstrates that electron spectra, laser field strength, and $\gamma$-photon spectra are unchanged for different pointing locations. Tilting the foam surface normal by $\Delta\phi=20^{\circ}$ from the laser axis also leaves the photon spectrum intact. This is a valuable feature since most laser facilities require off-normal target orientations to protect against back-reflections. Frequency-doubling can circumvent this requirement, but comes at the cost of total laser energy delivered on target. Hence, the foam's insensitivity to both pointing and steering is a desired feature for a robust $\gamma$-photon source.

The total photon yields from 2D simulations are approximate estimates, with full 3D effects needed to obtain a better absolute prediction. Nonetheless, the affordability of 2D simulations allows larger scans of target and laser parameters to obtain useful trends, as given in Fig.~\ref{fig:photons-scaling}. We start by scanning channel array width and foam filament spacing. Panels (a) and (b) present J/J conversion efficiency for $>1$ and $>100$ MeV photons, respectively. The lower-energy range favors dense foam targets with an average electron density approaching, but not exceeding, the relativistic critical density $\gamma n_c\approx140n_c$. In the provided scan, peak efficiency of $\eta_{\gamma}=15\%$ corresponds to $n_{e,avg}\approx55n_c$. The channel arrays also approach a similar efficiency for reduced filament spacing as the radiation mechanism shifts away from the $\gamma$-flash and toward the self-focusing regime.

Obtaining the highest-energy photons favors large filament spacings where the dominant process is the $\gamma$-flash. Trends in panel (b) indicate spacings larger than $w_0$ are favorable, whereas the relationship between $\eta_{\gamma}$ and channel length has a weaker scaling. The latter illustrates a complex balance between mean electron energy and the laser field strength reaching the reflector and driving the $\gamma$-flash.

We also scan laser energy for fixed $w_0=2.7$~\textmu m and $\tau_p=30$~fs across a range relevant for multi-PW laser facilities currently commissioned or under development, as shown in Fig.~\ref{fig:photons-scaling}(c). This includes ZEUS ($75$~J), ELI-NP ($200$~J), and NSF-OPAL ($500$~J) which all have a pulse duration of approximately $\tau_p=30$~fs at best compression. The NICS photon conversion efficiency demonstrates a near-linear scaling in this regime, reaching $31\%$ ($19\%$) for photons above 1~MeV (10~MeV) at NSF-OPAL conditions.

\section{Conclusion}

Additively-manufactured microstructured targets enable a new approach to designing $\gamma$-photon sources driven by nonlinear inverse Compton scattering (NICS). Channel arrays and staggered-filament foams offer two potential pathways for producing high-energy and high-flux $\gamma$-photon sources with designs that are compatible with modern two-photon polymerization (TPP) techniques. PIC simulations of these targets demonstrate distinct radiative patterns imprinted by the field configuration. Distilling these effects into a simple analytical model helps to simplify the target physics and aid design.

Channel arrays operate primarily in the $\gamma$-flash regime and are effective for producing collimated beams of high-energy photons. Finite wall thickness of the channels contribute to shot-to-shot fluctuations of photon yields due to laser pointing instabilities. The staggered-filament foam leverages self-focusing to drive strong magnetic and electric fields and promote NICS. While peak photon energies generated in the foam are lower than the channel arrays, the flux is significantly larger. The foams demonstrate excellent robustness to laser pointing instabilities and steering, serving as a major advantage for experimental deployment.

TPP targets are a promising approach to high-repetition rate experiments in high-field and HED science. Energetic and high-flux $\gamma$-photon sources will be important for studying fundamental strong-field physics like photonuclear effects and pair production, as well as supplying bright radiography sources to support HED experiments. Microstructured targets can play a pivotal role in controlling these $\gamma$-photon sources and maximizing the yields of laser-plasma interactions.

\section*{Acknowledgments}
This research is supported by DOE NNSA LRGF under the cooperative agreement DE-NA0003960 and the National Science Foundation (grant no. PHY-2611595). This work was performed under the auspices of the U.S. Department of Energy by Lawrence Livermore National Laboratory under contract DE-AC52-07NA27344. The authors would like to thank J. Ludwig and S. Wilks for their helpful feedback and discussions.

\bibliography{references}

\end{document}